\documentclass[twocolumn,aps,floatfix,superscriptaddress,pre,showpacs]{revtex4}
\usepackage{amsmath, amsthm, amssymb, graphicx}
\usepackage{color}

\input epsf.sty
\begin{document}

\title{Three-dimensional non-equilibrium Potts systems with magnetic friction}
\author{Linjun Li}
\affiliation{Department of Physics, Virginia Tech, Blacksburg, VA 24061-0435, USA}
\author{Michel Pleimling}
\affiliation{Department of Physics, Virginia Tech, Blacksburg, VA 24061-0435, USA}
\affiliation{Center for Soft Matter and Biological Physics, Virginia Tech, Blacksburg, VA 24061-0435, USA}
\affiliation{Academy of Integrated Science, Virginia Tech,
Blacksburg, VA 24061-0405, USA}
\date{\today}

\begin{abstract}
We study the non-equilibrium steady states that emerge when two interacting three-dimensional
Potts blocks slide on each other. 
As at equilibrium the Potts model exhibits different types of phase transitions
for different numbers $q$ of spin states, we consider the following three cases: $q=2$ (i.e. 
the Ising case), $q=3$, and $q=9$, which at equilibrium yield respectively a second order phase transition, 
a weak first order transition and a strong first order transition. In our study
we focus on the anisotropic character of the steady states that result from the relative motion and
discuss the change in finite-size signatures when changing the number $q$ of spin states. 
\end{abstract}


\maketitle

\section{Introduction}
Much of our knowledge on non-equilibrium steady states 
results from in-depth studies of transport models \cite{Cho11}, of driven systems \cite{Sch95},
as well as of reaction-diffusion systems \cite{Odo08}.
Our current understanding of non-equilibrium phase transitions has also profited greatly from
the investigation of model systems \cite{Hen08,Tau14}. Similar to the situation at equilibrium,
non-equilibrium phase transitions can either be continuous or discontinuous. Furthermore, cases
of strongly anisotropic phase transitions are also encountered far from equilibrium (see, e.g., the
driven lattice gas \cite{Sch95,Kat83}).

Spin models with magnetic friction and the related sheared spin models provide interesting classes 
of non-equilibrium systems that possess many intriguing properties 
\cite{Kad08,Fus08,Mag09,Mag09a,Huc09,Hil11,Igl11,Mag11,Mag11a,Ang12,Huc12,Mag13}.
The term magnetic friction is used to characterize the situation where spin correlations between
moving magnetic systems lead to energy dissipation. As a result the system settles into a non-equilibrium
steady state. Examples include a magnetic tip moving on a magnetic surface described as a classical
Heisenberg system \cite{Fus08,Mag09,Mag09a,Mag11,Mag11a,Mag13} as well as bulk spin systems moving relative
to each other \cite{Kad08,Hil11,Igl11,Ang12,Huc12}.
In \cite{Kad08} Kadau {\it et al} studied two coupled two-dimensional semi-infinite Ising models that slide on each other.
This sliding motion stabilizes the spin structure at the boundary, yielding an enhancement of the local
magnetization in cases where equal coupling strengths are considered everywhere in the system. 
Consequently, the boundary layers undergo a local phase transition at a temperature above the equilibrium bulk
critical temperature. This boundary phase transition temperature can be computed 
exactly for the two-dimensional Ising model in the limiting case of infinite
relative speed \cite{Huc09}. In \cite{Igl11} two-dimensional systems with magnetic friction
composed of Potts spins with $q$ states (the case $q=2$ being the Ising case) were considered.
This study revealed the existence of exotic non-equilibrium boundary phase transitions for large number of spin states $q$,
i.e. in situations where the equilibrium bulk system
undergoes a discontinuous transition. Indeed, depending on the strength of the boundary couplings between
the two sub-systems moving relatively to each other a change of the character of the non-equilibrium 
boundary phase transition is observed,
being continuous for weak boundary couplings and discontinuous when these couplings are strong. Hucht introduced
in \cite{Huc09} other Ising models with moving boundaries, including three-dimensional
geometries as well as sheared Ising systems. Later studies of some of these
Ising systems \cite{Ang12,Huc12} focused on the strongly anisotropic character of the non-equilibrium 
phase transitions encountered in these systems.

In the following we extend this line of research to coupled three-dimensional Potts blocks that slide on each other.
In three-dimensional bulk systems the Potts model displays different types
of equilibrium phase transitions as a function of the number of states $q$. We study in the following the cases
$q=2$ with a continuous equilibrium bulk phase transition, 
$q=3$ with a weak discontinuous phase transition in the bulk system, and $q=9$ where the
bulk transition is strongly discontinuous. Our aim is to develop a qualitative understanding of the non-equilibrium
steady states induced by the sliding of the blocks and to understand how the properties of the non-equilibrium 
boundary phase
transitions vary when changing the strength of the coupling between the blocks or the speed of the relative motion.

%
Our paper is organized in the following way. In the next Section we provide a more detailed discussion of
the studied geometry as well as of the local (boundary and line) quantities used to elucidate the properties
of our non-equilibrium systems. Section III is devoted to the numerical investigation of systems composed of two Potts
blocks that are in relative motion with respect to each other. 
Using local (boundary and line) quantities we investigate the magnetic properties of these driven systems as a 
function of relative speed as well as of the strength of the coupling between the two sub-systems. We discuss
the finite-size signatures in these anisotropic systems and elucidate how they change with the number $q$ of spin states.
We conclude in Section IV.

\section{Models}
We consider in this work three-dimensional models composed of
two $q$-state Potts systems that are coupled at their surfaces and
that move along their boundaries with a constant relative speed $v$. Each of the systems
is characterized by a lattice and a Hamiltonian of the form
\begin{equation}
{\mathcal H} = - J \sum\limits_{\langle \bf{r} , \bf{r}' \rangle} \, \delta \left(
S_{\bf{r}} - S_{\bf{r}'} \right)
\end{equation}
where the sum is over nearest neighbor lattice sites. The coupling constant $J$ is chosen to be
positive. $\delta$ is the Kronecker delta, with $\delta(x) = 1$ if $x =0$ and zero otherwise.
At every lattice site ${\bf r}$ we have a Potts spin $S_{\bf{r}}$ that takes on the values 
$S_{\bf{r}}=0, 1, \cdots , q-1$.

A Potts system with $q=2$ corresponds to the Ising model and exhibits in the 
thermodynamic limit a second order phase transition between a disordered high
temperature phase and an ordered low temperature phase. In a three-dimensional bulk system this transition
becomes a first order transition for $q \ge 3$. In our study we focus on three $q$ values,
$q=2$, 3, and 9, corresponding to a second order transition, a weak first order transition
and a strong first order transition, respectively.

Assuming that the surfaces are perpendicular to the $z$-direction and that the relative
motion is in the $y$-direction, we couple the two Potts systems through 
the time-dependent interaction term 
\begin{equation} \label{eq:V}
{\mathcal V}(t) = - J_b \sum\limits_{x_1,y_1}
\, \delta \left( S_{{\bf r}_1} - S_{{\bf r}_2(t)} \right)~,
\end{equation}
where ${\bf r}_1 = (x_1,y_1,z_1)$ is a lattice point in the surface layer $z_1$ of system 1, whereas
${\bf r}_2(t) = (x_2,y_2,z_2) = (x_1,y_1+ v t,z_1+1)$ is a surface point of system 2 located
above the site ${\bf r}_1$ but shifted in $y$-direction by the amount $v t$. 
This interaction term gives rise to magnetic friction and entails that the system settles into
a non-equilibrium steady state.
Besides varying
the dimensions of the sub-systems and the relative speed $v$, we will also consider different
coupling strength ratios $\kappa = J_b/J$ where $J_b$ is the strength of the couplings between the sub-systems,
whereas $J$ is the strength of the couplings within the sub-systems.

\begin{figure} [h]
\includegraphics[width=0.70\columnwidth,clip=true]{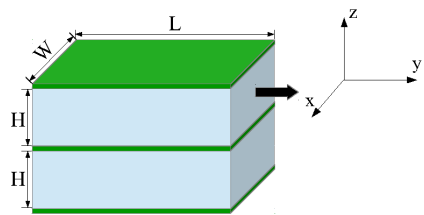}
\caption{\label{fig1} 
(Color online) 
Schematic picture of two identical sub-systems with relative motion in the $y$-direction.
The two sub-systems are blocks composed of $W \times L \times H$ spins, with periodic boundary conditions in
all three directions. Boundaries between the two sub-systems
are indicated by the green (dark) areas.
}
\end{figure}

The geometry discussed in this paper is shown in Fig. \ref{fig1}.
Our system is composed of two identical blocks 
where the upper block moves relative to the lower one. Typically the width $W$ and height $H$ vary between 
20 and 80 lattice sites. We investigate systems with length $L$ up to 240 sites
in order to check for anisotropy effects resulting
from the relative motion in $y$-direction.
As we are interested in the boundary properties, 
we use periodic boundary conditions in all three directions so
that every block experiences magnetic friction at two separate boundaries.

In order to elucidate the properties close to the boundary separating the
two sub-systems we focus on local quantities. Examples include the steady-state 
magnetization density in layer $z$ at temperature $T$
\begin{equation}
m(z,T) = \left( \frac{q \langle N_m(z,T) \rangle}{N(z)} -1\right) / \left( q -1 \right) 
\end{equation}
and the corresponding fluctuations around the mean layer magnetization density
\begin{equation}
\chi(z,T) = \frac{1}{k_B T N(z)} \left[ \langle N_m(z,T)^2 \rangle - \langle
N_m(z,T) \rangle^2 \right] ~.
\end{equation}
Here, $\langle N_m(z,T) \rangle$ is the average number of majority spins in layer $z$ at temperature $T$:
$\langle N_m(z,T) \rangle = \mbox{max} ( \langle N_0(z,T) \rangle, \cdots, 
\langle N_{q-1}(z,T) \rangle )$, where $\langle N_k(z,T) \rangle$ is the average number of spins
in state $k$ in layer $z$. The total number of spins in layer $z$ is denoted by $N(z)$,
with $N(z) = W \times L$ for the rectangular layers in the sub-systems shown in Fig. \ref{fig1}.
Another quantity of interest is the energy density in each layer $E(z,T)$ and the corresponding specific
heat $C(z,T) = dE(z,T)/dT$.
The boundary quantities are obtained by averaging over all equivalent boundary layers.
For the two-block system of Fig. \ref{fig1} we have four equivalent boundary layers, located
at $z=1$, $H$, $H+1$, and $2 H$, over which we can average in order to determine, for example, the mean
boundary magnetization density $m_b$ or the mean boundary specific heat $C_b$.

In order to probe for possible anisotropy effects resulting from the relative motion of the coupled
sub-systems, we also consider in the boundary layers the average line magnetizations in $x$- and $y$-directions, defined as
\begin{eqnarray}
m_x(y_0,T) & = & \left( \frac{q \langle N^x_m(y_0,T) \rangle}{N_x} -1\right) / \left( q -1 \right) \label{mx} \\
m_y(x_0,T) & = & \left( \frac{q \langle N^y_m(x_0,T) \rangle}{N_y} -1\right) / \left( q -1 \right) \label{my}
\end{eqnarray}
where $\langle N^x_m(y_0,T) \rangle$ (respectively $\langle N^y_m(x_0,T) \rangle$)
is the average number of majority spins in column $y_0$ (respectively row $x_0$) in the boundary layer at temperature $T$, 
whereas $N_x$ (respectively $N_y$) is the total number of spins in each column (respectively row).
For the rectangular layers of the sub-systems in Fig. \ref{fig1} we have that 
$N_x = W$ and $N_y= L$. As our systems are translationally invariant in $x$- and $y$-directions,
the choice of column $y_0$ and row $x_0$ is not important.

These magnetic properties are computed in Monte Carlo simulations where we follow previous work
and implement the relative motion between the two sub-systems by combining single spin updates
and shifts of a sub-system as a whole. For the single spin updates we use the standard
heat-bath algorithm. In order to simulate a system where one sub-system slides
with speed $v$ with respect to the other, we shift this sub-system by one lattice constant
after $N/v$ random single spin updates, where $N$ is the total number of spins in the system. 
One time step therefore consists of $N$ proposed
single spin updates and $v$ translations. 
Note that in the implementation we do not shift the sub-system that slides, but
instead only rewire the couplings at the boundary, as this involves much fewer computational operations.

\section{Sliding Potts blocks}

In this Section we study the magnetic properties of three-dimensional Potts spin blocks sliding
past each other. Results for these systems are scarce, and the only previous result that directly
relates to our study is the calculation of the shift of the critical temperature for the case of
two Ising blocks with couplings of only one strength (i.e. $J_b=J=1$)
moving with infinite relative speed. Indeed, in that case the critical temperature of
the non-equilibrium system can be expressed as a function of the zero-field equilibrium susceptibility \cite{Huc09}.
From the eighth-order high temperature series for the equilibrium susceptibility one
finds in Potts units the critical temperature $T_c=2.40(5)$ \cite{Huc09}, substantially larger than the critical temperature 
$T_c=2.256$ of the three-dimensional equilibrium Ising model. 

On general grounds, the phase transition in our system is expected to be strongly anisotropic, similar
to what is observed in related cases \cite{Ang12,Huc12}. In what follows, we indeed discuss the anisotropic
properties close to the phase transition, and this in cases where the bulk transition is either continuous or
discontinuous. We limit ourselves to a qualitative discussion, leaving a more quantitative study (which
requires for the cases of large number of states $q$ resources not currently available to us) for a later time.

\begin{figure} [h]
\includegraphics[width=0.95\columnwidth]{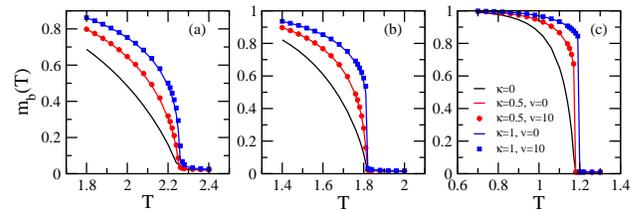}
\caption{\label{fig2} (Color online)
Boundary magnetization density as a function of temperature $T$
for two Potts blocks that are either both at rest ($v=0$) or where one of the blocks
is moving with respect to the other with speed $v=10$. Data for different small values
of the coupling strength ratio $\kappa$ are shown. For $\kappa =0$ the two blocks are 
uncoupled, whereas for $\kappa =1$ the couplings at the boundary have the same strength
as the couplings inside the bulk. The number of states are (a) $q=2$, (b) $q=3$, and (c)
$q=9$. Every block is composed of $80 \times 80 \times 80$ spins.
The data result from averaging over at least ten independent runs, and error bars are smaller than the 
symbol sizes.
}
\end{figure}

Fig. \ref{fig2} shows for systems composed of blocks with $80 \times 80 \times 80$ spins
the temperature-dependent boundary magnetization density $m_b(T)$ for a variety of cases
with vanishing or small coupling strength ratios $\kappa$. The full lines correspond to equilibrium 
situations, whereas the symbols give values in non-equilibrium steady states. For $\kappa =0$ (black lines) the two sub-systems are
uncoupled, and $m_b$ is then the surface magnetization of an equilibrium system with open boundary conditions.
For all values of $q$ the surface magnetization vanishes continuously with increasing temperature, and this even so for $q \ge 3$ the
equilibrium bulk transition is discontinuous. This surface-induced disordering effect is well known for
systems with free surfaces where the bulk undergoes a discontinuous transition \cite{Lip82,Lip83,Sch96,Igl99,Haa00,Li13}.
For $\kappa =1$ and $v=0$ we recover the bulk equilibrium system with a discontinuous phase transition for $q \ge 3$,
i.e. there is a critical value of $\kappa$ between 0 and 1 
at which the character of the equilibrium boundary transition changes, see Fig. \ref{fig2}.

Only minor differences between the equilibrium and non-equilibrium cases with $v=10$ can be seen in Fig. \ref{fig2} for $0 < 
\kappa \le 1$.
A closer look reveals for $q=2$ and $\kappa =1$ that the symbols lie systematically above the equilibrium results, in agreement
with the predicted shift of the critical temperature \cite{Huc09}. The same holds true for $q=3$, whereas for $q=9$
equilibrium and non-equilibrium data are identical within error bars. 
For the Ising case the data are compatible with the expected shift of $T_c$ \cite{Huc09}, but seem to indicate
a much smaller increase than that obtained by Hucht for the case $v = \infty$
from (admittedly rather short) high temperature series for the equilibrium susceptibility. The reader, however,
should note that a square boundary layer is not the most appropriate geometry close to the phase
transition. As we argue below,
anisotropic samples are much better suited in order to obtain quantitatively correct data in vicinity of the strongly
anisotropic phase transition.

In a previous study of the two-dimensional Potts system with $q=9$ states where two halves of the system slide on top
of each other \cite{Igl11} a change of the character of the boundary transition as a function of $\kappa$ was also observed:
for small values of $\kappa$ the boundary phase transition is continuous and takes place at the bulk transition
temperature, whereas for large values of $\kappa$ the transition is discontinuous and the transition temperature
is shifted to values larger than the bulk transition temperature. However, in this case the ordering of the boundary 
(which is a one-dimensional object) at a temperature above the bulk transition temperature
is a purely non-equilibrium effect as in equilibrium a one-dimensional spin system with
short-range interactions does not support long-range order.

\begin{figure} [h]
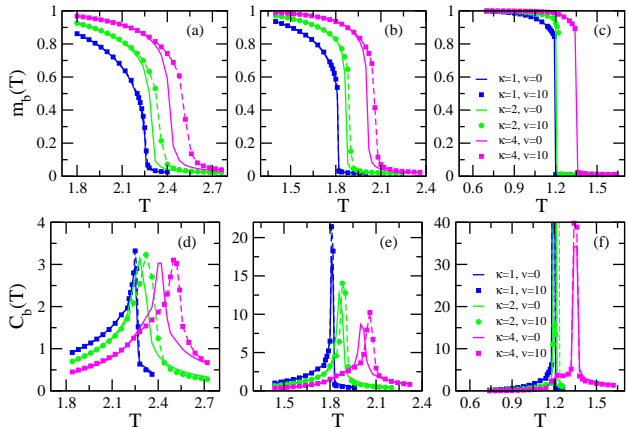

\includegraphics[width=0.95\columnwidth]{figure3a.eps}\\
\includegraphics[width=0.95\columnwidth]{figure3b.eps}
\caption{\label{fig3} (Color online)
(a)-(c) Boundary magnetization density $m_b$ and (d)-(f) boundary specific heat $C_b$ as a function of temperature $T$
for two Potts blocks that are either both at rest ($v=0$) or where one of the blocks
is moving with respect to the other with speed $v=10$. Data for different large values
of the coupling strength ratio $\kappa$ are shown. The number of states are (a,d) $q=2$, (b,e) $q=3$, and (c,f)
$q=9$. The relative motion stabilizes the ordering of the boundary
which results in an additional shift of the local transition temperature. This is clearly visible in the data for $q=2$ and $q=3$.
Every block is composed of $80 \times 80 \times 80$ spins.
The data result from averaging over at least ten independent runs, and error bars are smaller than the
symbol sizes.
}
\end{figure}

When further increasing the strength of the coupling between the sub-systems, the trends already visible in Fig. \ref{fig2} persist
and become very pronounced. This is illustrated in Fig. \ref{fig3} through the temperature dependence of the 
boundary magnetization density $m_b(T)$ as well as of the boundary specific heat $C_b(T)$. 
We note that for all values of $q$ and $\kappa > 1$ the
phase transition temperature is larger than that of the perfect equilibrium bulk system with $v=0$ and $\kappa = 1$. 
This shift is readily understood for the equilibrium case (full lines) as the boundary region with a strong coupling between
the sub-systems behaves like a two-dimensional object that orders at a higher temperature than the bulk.
The sliding motion further enhances this tendency for increased ordering, and for $q=2$ and $q=3$ an additional increase
of the boundary transition temperature, that results from the motion, is observed (see the symbols and dashed lines
in Fig. \ref{fig3}). For very large values of $q$, see the case $q=9$ in Fig. \ref{fig3}c and \ref{fig3}f, 
the relative motion only results in very minor changes with respect to the equilibrium situation.

\begin{figure} [h]
\includegraphics[width=0.95\columnwidth]{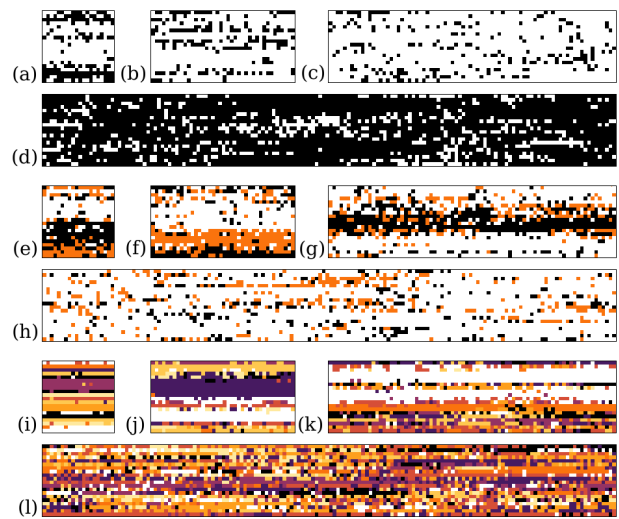}
\caption{\label{fig4} (Color online)
Snapshots of one of the boundaries of a system formed by two blocks with $20 \times L \times 10$
spins. The speed is $v=10$ and the ratio of coupling strengths is $\kappa = 9$. (a)-(d) $q=2$ 
and $T=3$, (e)-(h) $q=3$ and $T=2.45$, (i)-(l) $q=9$ and $T=1.55$. The length of the sample is
(a) $L=20$, (b) $L=40$, (c) $L=80$, and (d) $L=160$, and similarly for the other two values of $q$.
The different colors correspond to the different states of the spins.
}
\end{figure}

\begin{figure} [h]
\includegraphics[width=0.60\columnwidth]{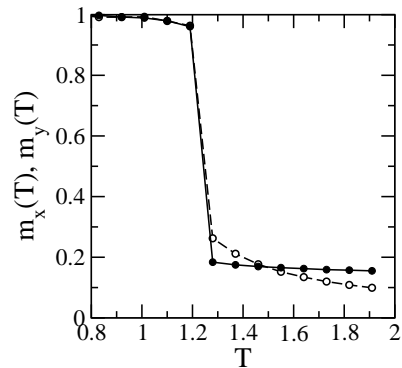}
\caption{\label{fig5} 
Line magnetization densities in $x$ (filled squares) and $y$ (open circles) directions for a system
with $q=9$, $\kappa =9$, and $v=10$ composed of blocks containing $20 \times 160 \times 10$ spins. A typical 
spin configuration for that system at $T = 1.55$ is shown in Fig. \ref{fig4}(l).
}
\end{figure}

Looking at boundary quantities like those in Fig. \ref{fig2} and \ref{fig3} does not provide a comprehensive
view of our systems, as they do not reveal the anisotropy effects induced by the relative motion of the blocks.
Fig. \ref{fig4} shows some typical spin configurations in the boundary layer
for systems with large boundary couplings and different 
aspect ratios, taken at temperatures close to the boundary phase transition. 
We focus in the following on the $q=9$ states case shown in Fig. \ref{fig4}(i)$-$(l), but the same effects
are observed for other number of spin states, see the figure. Starting with a square boundary in panel (i), we double from panel
to panel the length of the boundary in direction of the relative motion
until the aspect ratio is 8 for panel (l). The motion of the blocks induces
additional correlations in the sliding direction, and one therefore expects anisotropy effects to show up
as direction-dependent correlation lengths and, in the ordered phase, anisotropically shaped ordered domains.
For the example shown in panel (i) to (l), the anisotropy effects in
systems with small aspect ratios take the form of almost completely ordered lines
in the direction of motion (horizontal or $y$-direction), whereas in the direction perpendicular to
the motion (vertical or $x$-direction) the spins are much more disordered. Even so we are close to the phase transition,
the smallness of the horizontal dimension yields as an artifact a very large line magnetization. Increasing
the length of the system in that direction allows to capture better and better the fluctuations and yields
for the largest length shown in panel (l) configurations with a comparable level of order in both directions.
This is illustrated in Fig. \ref{fig5} where we compare the line magnetization densities in the different directions
for the system with blocks containing $20 \times 160 \times 10$ spins.

\begin{figure} [h]
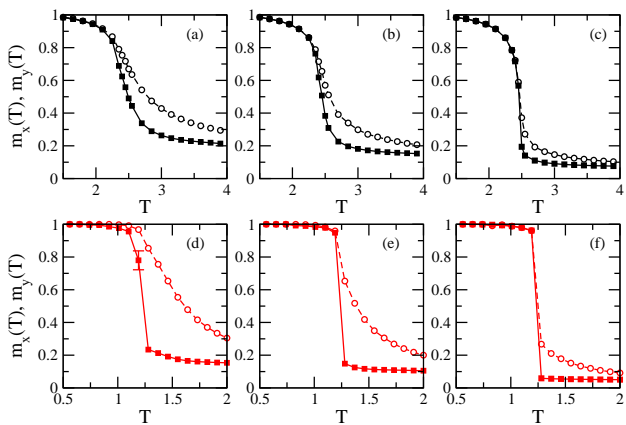

\includegraphics[width=0.95\columnwidth]{figure6a.eps}\\
\includegraphics[width=0.95\columnwidth]{figure6b.eps}
\caption{\label{fig6} (Color online)
Line magnetization densities in $x$ (filled squares) and $y$ (open circles) directions for $\kappa = 9$. 
(a)-(c) $q=2$ and (d)-(f) $q=9$. The sizes of the blocks, which move with relative speed $v=10$,
are $L \times L \times 20$, with
(a,d) $L=20$, (b,e) $L=40$, and (c,f) $L=160$. The data result from averaging over at least 
10 independent runs. Error bars are only shown when the error is larger than the symbol size.
}
\end{figure}

Fig.\ \ref{fig6} shows some quantitative data for the line magnetization densities $m_x$ and $m_y$
(see equations (\ref{mx}) and (\ref{my})) for $q=2$ and $q=9$, with 
$\kappa = 9$. In order to compare finite-size effects, we consider square boundaries with $L \times L$ spins,
$L$ ranging from 20 to 160. 
The data for $q=2$ in the first row show the expected finite-size behavior of the line magnetization
close to an anisotropic critical point: in the direction of motion fluctuations are more strongly constrained,
which yields a higher level of ordering as witnessed by the larger line magnetization density $m_y$ (open symbols).
Increasing the system size allows to capture better and better the fluctuations, and the two densities get
increasingly comparable. An interesting additional effect shows up when considering the system with a discontinuous
bulk transition as it is the case for $q=9$, see second row of Fig.\ \ref{fig6}. As seen in Fig. \ref{fig3}c,
the boundary transition is also discontinuous in that case for large values of $\kappa$, as evidenced by the discontinuity
in the boundary magnetization. However, for the smaller systems only the line magnetization $m_x$ in direction perpendicular
to the motion (filled squares) displays a discontinuous character. The line magnetization $m_y$ in direction of the motion
shows a smooth behavior, see panel (d), reminiscent of that observed in panel (a) for $q=2$ where the boundary 
transition is continuous. It is only for larger systems that also $m_y$ starts to reveal a large jump indicating
the discontinuous character of the transition in the thermodynamic limit.

\begin{figure} [h]
\includegraphics[width=0.95\columnwidth]{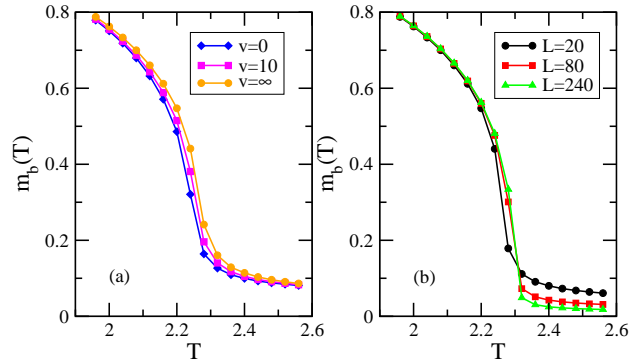}
\caption{\label{fig7} (Color online)
Boundary magnetization densities for the Ising model as a function of temperature. (a) Shift 
of the boundary magnetization density as a function of the speed $v$ 
in systems composed of blocks containing $20 \times 20 \times 20$ spins. (b) Boundary magnetization densities
for anisotropically shaped samples with $40 \times L \times 20$ spins in each block that move
with relative speed $v = \infty$. Based on all our data, we estimate the critical temperature to be $T_c = 2.34(2)$.
}
\end{figure}

As discussed previously, the data shown in Fig. \ref{fig2} for $\kappa =1$ seem to indicate for the Ising case 
a rather small shift of the critical temperature compared to the equilibrium case. We have another look at 
this in Fig. \ref{fig7} where we consider samples moving with different speeds as well as different anisotropic shapes. 
We here consider also the case $v = \infty$
where for the update of a boundary spin, located at, say, the top of the lower block at site ${\bf r}_1 = (x_1,y_1,H)$, we connect
this spin via the coupling term (\ref{eq:V}) to a randomly selected spin with the same $x$-coordinate
but located in the neighboring boundary layer \cite{Huc09}, i.e. this second spin has the coordinates ${\bf r}_2 = (x_1,y_2,H+1)$
with $1 \leq y_2 \leq L$. Fig. \ref{fig7}a illustrates the shift of the magnetization densities due to the
relative motion for blocks composed of $20 \times 20 \times 20$ spins.
As shown in Fig. \ref{fig7}b, strong finite-size effects
do not allow to obtain reliable estimates of the critical temperature for small
values of the aspect ratio. For larger 
aspect ratios these effects vanish. Based on our data we obtain the estimate $T_c = 2.34(2)$ for the Ising model
with $v = \infty$. We have a rather good agreement
with the estimate $T_c = 2.40(5)$ obtained by Hucht \cite{Huc09},
especially when taking into account that for the equilibrium three-dimensional Ising model
the known series for the zero-field susceptibility, used in \cite{Huc09}, are rather short (only up to eighth order).

As already mentioned at the beginning of this Section, our main interest here is to understand qualitatively
the characteristic features of the boundary transition in the Potts model and to compare cases where the bulk
transition is continuous with those where this transition is discontinuous. Studying in detail the properties
of the strongly anisotropic non-equilibrium critical point that shows up in the former case is beyond the current
work and would need additional extensive numerical simulations. In any case, we do not anticipate a 
behavior that differs markedly from that revealed in related Ising systems in two space dimensions \cite{Ang12,Huc12}.

\section{Conclusion}
In this work we studied the magnetic properties of three-dimensional Potts systems where two coupled blocks
are shifted against each other with some speed $v$. Because of this shift, the system settles into
a non-equilibrium steady state. Increasing the temperature then yields a non-equilibrium phase transition
between an ordered low temperature phase and a disordered high temperature phase.

Depending on the number of spin states $q$, the temperature-driven phase transition in the
equilibrium three-dimensional Potts system can be either continuous (for $q < 3$) or discontinuous
(for $ q \geq 3$). In our investigation we considered the different situations $q=2$ (Ising case),
$q=3$ (weakly discontinuous) and $q=9$ (strongly discontinuous). Our study revealed some common features
that are independent of the value of $q$, but also showed the existence of marked differences between
the different cases. Whereas for small numbers of spin states ($q=2$ and $q=3$) the transition temperature 
between the disordered and ordered phases is shifted to higher values due to the relative motion of
the blocks, no such shift is observed for
large values of $q$. On the other hand, intriguing finite-size effects are encountered for large $q$ values
where in smaller samples the discontinuous character of the boundary phase transition is not showing up in
the seemingly continuous variation of the line magnetization in the direction of the relative motion.

Common to all the cases is the emergence of additional correlations in the direction of relative motion.
As a result the phase transition temperature is shifted to higher values in cases where the coupling between
the sub-systems is not too weak. The value of the shift depends on the value of the relative speed. Another
consequence of these additional correlations is the strongly anisotropic character of the phase transition.
In computer simulations this entails rather complicated finite-size effects that necessitate anisotropically
shaped samples in order to capture the typical fluctuations close to the phase transition. These finite-size
effects show up in different forms, depending on the value of $q$. For example, for large $q$ and small
lengths in the direction of the motion the line magnetization density $m_y$, which results
from averaging along the direction of motion, displays a smooth behavior. Only after increasing the size
of the sample in that direction (i.e. increasing the aspect ratio) does the discontinuous character of
the transition show up also in this quantity.

The present study is clearly not exhaustive and many possible future research directions can be envisioned.
For example, interesting open questions remain for the equilibrium case $v=0$. Indeed for coupling strength
ratios $\kappa \ne 1$ we are dealing with a three-dimensional bulk system with a planar defect. Defects have
been shown to yield intriguing local critical phenomena in bulk systems undergoing a phase transition (see
\cite{Igl93} for a review of some of these phenomena). However, most of these studies have been restricted
to two-dimensional systems (see \cite{Bar79,Hil81,Igl90,Igl93a} for some examples), 
where analytical approaches are possible, whereas in three dimensions not much
is known beyond mean-field level considerations. This investigation of the static local critical properties
could be augmented by an investigation of relaxation processes, similarly to what has been done previously
for two-dimensional systems with defects \cite{Ple05}. Similar issues can be studied for non-equilibrium
cases with $v>0$. However, in that situation we expect as further complication to have to deal with a strongly
anisotropic critical behavior with direction dependent correlation length exponents, similar to what has been
observed in \cite{Ang12} for the special case of two planar Ising models that are moved relative to each other.

\begin{acknowledgments}
This work is supported by the US National
Science Foundation through grant DMR-1205309. We thank Alfred Hucht for useful discussions.
\end{acknowledgments}

\end{document}